\documentclass{article}
\usepackage{graphicx}
\usepackage{hyperref}
\usepackage{natbib}

\title{The Cyber Immune System: Harnessing Adversarial Forces for Security Resilience}
\author{
  Dr. Krti Tallam \\
  EECS, University of California at Berkeley \\}
\date{\today}

\begin{document}

\maketitle

\begin{abstract}
Both parasites in biological systems and adversarial forces in cybersecurity are often perceived as threats—disruptive elements that must be eliminated. However, these entities play a critical role in revealing systemic weaknesses, driving adaptation, and ultimately strengthening resilience. This paper draws from environmental epidemiology and cybersecurity to reframe parasites and cyber exploiters as essential stress-testers of complex systems, exposing hidden vulnerabilities and pushing defensive innovations forward. By examining how biological and digital systems evolve in response to persistent threats, we highlight the necessity of adversarial engagement in fortifying security frameworks. The recent breach of the DOGE website serves as a timely case study, illustrating how adversarial forces—whether biological or digital—compel systems to reassess and reinforce their defenses.
\end{abstract}

\section{Introduction}
In any complex system, certain forces operate at the margins—often dismissed as threats or disruptions. Yet, these forces can act as stressors that test, expose, and ultimately strengthen a system’s resilience. In biological ecosystems, parasites persist by infiltrating hosts, revealing evolutionary vulnerabilities, and driving adaptive responses \citep{dobson1989population, anderson1991infectious}. In digital ecosystems, adversarial agents serve a similar function, identifying weaknesses in cybersecurity infrastructures and compelling innovations in defense strategies \citep{schneier2000secrets, mitnick2003art}. 

The COVID-19 pandemic provided a stark example of how unseen forces can rapidly exploit systemic weaknesses. The virus did not just challenge immune systems; it exposed vulnerabilities in global supply chains, healthcare infrastructure, and crisis response mechanisms \citep{lakner2021measuring, hsiang2020effect}. Much like a sophisticated cyberattack, it infiltrated networks, spread unpredictably, and necessitated a reevaluation of preparedness strategies \citep{lu2020artificial, chang2020covid}. The lag in detection, the rapid mutations, and the need for adaptive defenses mirror the challenges of modern cybersecurity, where adversarial actors continually refine their methods in response to evolving countermeasures \citep{biggio2018wild}. 

Adversarial forces—whether biological or digital—often emerge in a system’s blind spots. Their presence raises fundamental questions: How do systems recognize vulnerabilities before they are exploited? What mechanisms enable adaptation, and which weaknesses remain undetected until tested? Most critically, can these disruptive forces be leveraged as tools for strengthening resilience rather than being framed solely as threats? 

Parasites have long been studied for their ability to manipulate hosts, yet their presence has also driven evolutionary progress. Immune systems, behavioral defenses, and ecological stability have all been shaped by selective pressures imposed by parasitic organisms \citep{morens2020emerging, schmid2014antigenic}. Similarly, in cybersecurity, the ongoing tension between adversarial actors and defenders has led to the development of stronger encryption protocols, dynamic authentication mechanisms, and adaptive risk assessment models \citep{shostack2014threatmodeling, barabasi2016network}. 

By reconsidering these disruptive agents as catalysts for system evolution, we recognize that resilience is not built through static defenses but through continuous adaptation. The evolutionary interplay between parasites and hosts has driven increasingly sophisticated immune responses, just as the interplay between cyber threats and security research has resulted in more robust digital protections. Both domains, seemingly distinct, follow a common principle: \textbf{adversarial interactions expose systemic weaknesses, and through iterative responses, systems evolve toward greater resilience.}

This paper explores how insights from parasitology can inform cybersecurity frameworks. How do parasitic adaptation strategies mirror those of digital intrusions? What lessons from biological resilience can be applied to securing digital ecosystems? By examining these questions, we challenge conventional notions of security—not as the eradication of threats, but as a continuous, adversarially driven process of reinforcement and refinement. 

\textbf{Rather than viewing adversarial actors as mere disruptors, we argue that their presence is fundamental to system resilience.} This perspective shifts cybersecurity away from purely defensive paradigms and toward a model that actively integrates adversarial interactions as an essential driver of innovation and long-term security.

\section{Background and Literature Review}
The study of systemic vulnerabilities, whether in biological or digital ecosystems, has been an area of interdisciplinary research spanning evolutionary biology, cybersecurity, epidemiology, and risk assessment. Understanding how systems respond to persistent threats requires insights from ecological stability, immunology, network security, and complexity science.

\subsection{Biological Perspectives on Systemic Threats}
In biological systems, parasitology provides a foundational framework for studying host-pathogen interactions, adaptation mechanisms, and co-evolutionary pressures. The seminal work of Anderson and May \citep{anderson1991infectious} models host-parasite population dynamics, illustrating how persistent infections drive the evolution of host immune responses. Similarly, epidemiological research has demonstrated how infectious diseases exert selective pressure on populations, leading to genetic adaptations that enhance resistance \citep{dobson1989population}. The concept of \textbf{immunological memory}—where prior exposure to pathogens enhances future immune responses—mirrors the iterative security updates and defenses developed in digital systems to mitigate evolving cyber threats.

Zoonotic disease outbreaks, such as HIV, SARS, and COVID-19, further underscore how ecological disruptions and human-mediated environmental changes can amplify systemic vulnerabilities \citep{morens2020emerging}. The principles used to track and contain pandemics bear strong methodological similarities to those used in cybersecurity: just as epidemiologists model disease spread to implement containment strategies, cybersecurity analysts monitor digital threat landscapes, identifying vulnerabilities before they lead to cascading failures.

\subsection{Cybersecurity and Adversarial Threats}
Cybersecurity has evolved alongside technological advancements, with adversarial actors continuously identifying and exploiting system weaknesses. Foundational works such as Schneier’s analysis of security engineering \citep{schneier2000secrets} and Mitnick’s research on social engineering \citep{mitnick2003art} reveal that vulnerabilities in digital systems are often not just technical but also deeply human. Reports from CISA and NIST \citep{cisa2022supply} highlight the growing prominence of supply chain attacks, where adversaries exploit interconnected dependencies—an approach not unlike how parasites leverage ecological networks to propagate across hosts.

As cyber threats become more sophisticated, the role of \textbf{adversarial learning} in security has grown. AI-driven attacks in adversarial machine learning \citep{biggio2018wild} demonstrate that subtle perturbations can manipulate model outputs, exposing hidden weaknesses in ways reminiscent of how certain parasites evade immune detection through molecular mimicry \citep{schmid2014antigenic}. This raises urgent questions about robustness in AI security, particularly as autonomous systems take on a larger role in threat detection and mitigation.

\subsection{Complexity Science and Resilience Strategies}
Complexity theory provides a valuable framework for understanding adversarial interactions across biological and digital domains. Taleb’s concept of antifragility \citep{taleb2012antifragile} argues that systems gain resilience through exposure to stressors—a principle that underlies both host immune defenses and cybersecurity resilience strategies. Likewise, network theory and research on cascading failures highlight how small vulnerabilities, if unaddressed, can propagate throughout an entire system, whether in biological immunity or digital infrastructure \citep{barabasi2016network}.

By integrating insights from these disciplines, this paper frames adversarial agents—whether biological parasites or digital exploiters—not as anomalies to be eliminated but as necessary components of system evolution. \textbf{Security is not about achieving perfect immunity; it is about developing adaptive, resilient structures that continuously evolve in response to disruption.} This perspective challenges conventional security paradigms, advocating for proactive engagement with threats rather than reactionary containment alone.

\section{Parasitology and Cybersecurity: A Comparative Analysis}
\subsection{Parasites: Adaptive Exploiters in Biological Systems}
Parasites have evolved highly specialized mechanisms to infiltrate, exploit, and persist within their hosts. These strategies range from biochemical mimicry to immune suppression, allowing them to manipulate host physiology and behavior \citep{schmid2014antigenic}. For example, the protozoan parasite \textit{Toxoplasma gondii} can alter rodent behavior, increasing the likelihood of predation by felines, its definitive host \citep{dobson1989population}. This behavioral manipulation serves as an adaptive strategy for transmission, highlighting how parasites evolve to exploit host vulnerabilities.

Beyond individual host manipulation, parasites can destabilize entire ecosystems by exposing systemic weaknesses. The chytrid fungus, which has devastated global amphibian populations, demonstrates how a single adaptive pathogen can disrupt biodiversity on a large scale \citep{fisher2009emerging}. Similarly, vulnerabilities in digital networks can be exploited by malicious software or automated cyber threats, propagating rapidly across interconnected systems before adequate defenses can be deployed \citep{cisa2022supply}.

Parasites and their hosts are engaged in an ongoing evolutionary arms race, where hosts develop immune responses while parasites continuously adapt to evade them \citep{anderson1991infectious}. This dynamic mirrors cybersecurity’s continuous cycle of attack and defense, where adversarial actors identify new vulnerabilities, prompting security researchers to refine protective measures \citep{schneier2000secrets}. Understanding these co-evolutionary dynamics provides insight into designing adaptive cybersecurity models that evolve in response to emerging threats.

\subsection{Adversarial Agents in Digital Security}
Adversarial actors in cybersecurity, much like biological stressors, exploit weaknesses in digital systems to achieve their objectives. These actors can be categorized into several groups:
\begin{itemize}
    \item \textbf{Malicious adversaries}: Individuals or groups that exploit security flaws for financial, political, or disruptive gain, often causing harm to individuals or institutions \citep{mitnick2003art}.
    \item \textbf{Ethical security testers}: Professionals who work within structured environments to identify and remediate vulnerabilities before they can be exploited maliciously \citep{harris2020ethicalhacking}.
    \item \textbf{Unstructured security researchers}: Individuals who operate in a legal and ethical gray area, often disclosing vulnerabilities but without prior authorization \citep{holt2012hackers}.
\end{itemize}

The methodologies used in cybersecurity often parallel biological adaptation strategies. Phishing attacks rely on social engineering to deceive users into providing access—an approach that exploits behavioral vulnerabilities, much like parasites manipulate host behavior to enhance transmission \citep{ferguson2018socialengineering}. Similarly, malware evasion techniques, such as trojans masquerading as legitimate software, resemble how some parasites mimic host cells to avoid detection by the immune system \citep{biggio2018wild}.

Importantly, structured adversarial testing, such as penetration testing and red teaming, plays a role analogous to controlled immunization strategies in biology. Just as exposure to weakened pathogens trains the immune system to recognize and neutralize threats, proactive security assessments help organizations identify and address vulnerabilities before they are exploited in real-world scenarios \citep{shostack2014threatmodeling}. 

Ultimately, both biological and digital systems must balance adaptation and resilience in the face of persistent adversarial pressures. By studying how hosts and pathogens co-evolve, cybersecurity professionals can develop proactive defense strategies, leveraging techniques such as adaptive threat modeling, continuous monitoring, and real-time anomaly detection \citep{barabasi2016network}. The challenge lies not in eliminating adversarial forces entirely, but in designing resilient systems that can withstand and adapt to emerging threats.

\begin{figure}
    \centering
    \includegraphics[width=1\linewidth]{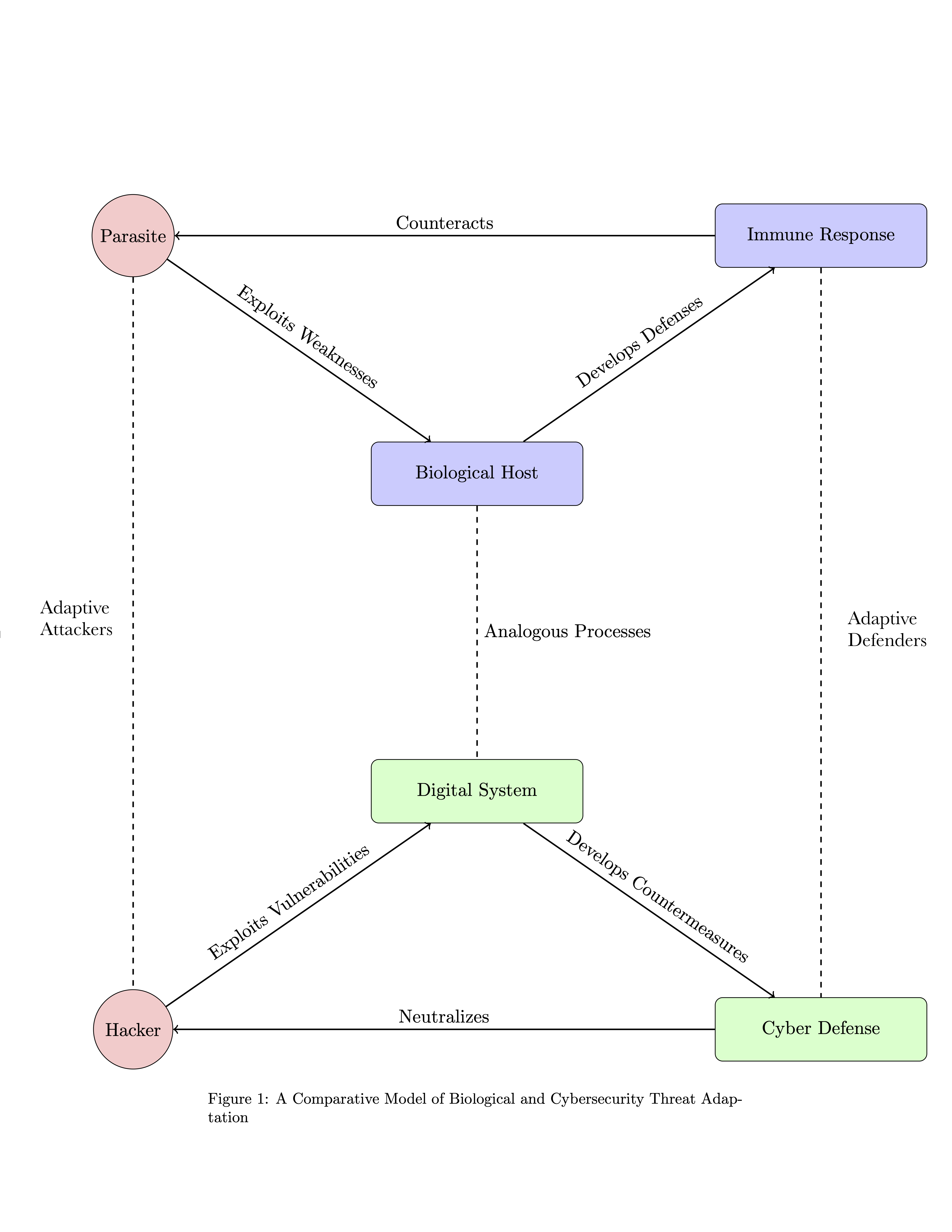}
    \label{fig:Parasites_Hackers.png}
\end{figure}

\section{Lessons from Adversarial Stressors in Biological and Digital Systems}
\subsection{Resilience and Adaptation}
Resilience in both biological and digital systems does not arise from stability but from persistent adaptation in response to external pressures. In biological ecosystems, hosts evolve resistance against parasites through genetic mutations, immune system enhancements, and behavioral modifications \citep{anderson1991infectious}. Likewise, cybersecurity frameworks continuously refine their encryption standards, anomaly detection models, and intrusion prevention mechanisms to counteract emerging cyber threats \citep{shostack2014threatmodeling}.

The interplay between adversarial actors and defense mechanisms in both domains follows a coevolutionary arms race: parasites adapt to bypass immune defenses, while hosts develop increasingly sophisticated responses \citep{schneier2000secrets}. In cybersecurity, the same dynamic is evident, as adversarial agents develop new attack strategies, prompting defenders to create more resilient security architectures. This iterative cycle of adaptation is exemplified in Red Team vs. Blue Team exercises, where security professionals simulate adversarial attacks to refine defensive strategies—an approach analogous to the way natural selection continuously optimizes immune responses \citep{harris2020ethicalhacking}.

Beyond direct competition, certain adversarial forces can also contribute positively to system stability. In biological ecosystems, some parasites evolve toward less harmful relationships with their hosts, ensuring their own long-term survival \citep{dobson1989population}. Similarly, ethical hackers and security researchers act within digital ecosystems to identify and remediate vulnerabilities before they can be exploited maliciously \citep{holt2012hackers}. This suggests that effective security frameworks should not seek to eliminate adversarial forces entirely but should instead embrace continuous stress-testing and resilience-building as integral components of systemic defense.

\subsection{Weak Points and Fortifications}
Adversarial forces do more than exploit weaknesses—they reveal them, compelling systems to adapt and fortify their defenses. Just as parasitic pressures have driven the evolution of immune complexity \citep{schmid2014antigenic}, persistent cybersecurity threats accelerate the refinement of security infrastructure. 

In biology, organisms that fail to recognize and neutralize threats succumb to infections, while those that develop robust immune responses persist and pass on their adaptive advantages. This same principle applies in cybersecurity: the discovery of exploitable vulnerabilities forces the development of more secure coding practices, stronger encryption standards, and improved security architectures \citep{barabasi2016network}.

A real-world example of this dynamic is the DOGE website hack, in which attackers exploited overlooked security gaps to gain unauthorized access. This breach underscored the importance of proactive security measures, such as rigorous security audits, multi-factor authentication, and continuous monitoring, as essential components of digital resilience \citep{cisa2022supply}. Just as disease outbreaks expose weaknesses in public health systems, cyberattacks force organizations to reevaluate and strengthen their security postures.

Furthermore, research in epidemiology and immunology suggests that controlled exposure to weaker threats—such as through vaccination—can prime immune systems for stronger responses against more virulent pathogens \citep{morens2020emerging}. This principle is reflected in cybersecurity through penetration testing and ethical hacking, where simulated attacks proactively identify and remediate weaknesses before malicious actors can exploit them \citep{biggio2018wild}. By applying these lessons, security frameworks can shift from reactive mitigation to proactive resilience-building, ensuring that systems are continuously stress-tested and iteratively improved.

Ultimately, both biological and digital systems do not remain resilient by eliminating threats entirely, but by engaging with adversarial forces as necessary stressors that drive continual adaptation. Recognizing adversarial interactions as integral to resilience shifts security away from static defenses toward an evolving, self-reinforcing model that thrives in dynamic threat environments.

\section{Reframing the Narrative}
Adversarial forces, whether in biological or digital systems, are often viewed as purely destructive. However, their presence plays a crucial role in strengthening system resilience. Just as controlled exposure to pathogens in vaccines enhances immune defenses, structured adversarial testing—through ethical hacking and penetration testing—fortifies cybersecurity systems by revealing vulnerabilities before they can be exploited.

Historically, attempts to eliminate all perceived threats have often resulted in unintended consequences. Over-sterilization in medical environments, for instance, has been linked to weaker immune responses and increased susceptibility to disease \citep{schmid2014antigenic}. Similarly, rigid cybersecurity strategies that fail to anticipate adaptive adversaries have left organizations vulnerable to systemic risks, as demonstrated by supply chain attacks and zero-day exploits \citep{cisa2022supply}. Resilience is not built through static defenses but through continuous adaptation in response to stressors.

Recognizing the constructive role of adversarial interactions allows for the design of more robust and adaptable security frameworks. In both immunology and cybersecurity, the concept of \textbf{live-fire exercises} demonstrates that controlled exposure to threats leads to better preparedness. In biological systems, vaccines introduce weakened or inactive forms of a virus to train the immune system without causing illness \citep{morens2020emerging}. In cybersecurity, penetration testing and ethical hacking simulate real-world cyberattacks, enabling security teams to refine defenses without suffering catastrophic breaches \citep{shostack2014threatmodeling}. Both approaches acknowledge that resilience is not achieved by avoiding threats entirely but by engaging with them in controlled, strategic ways.

Policy frameworks must strike a balance between mitigating risks and allowing ethical hacking and security research to inform systemic fortifications. Overregulating security testing could stifle innovation and leave vulnerabilities undiscovered, just as indiscriminate anti-parasitic interventions have sometimes led to ecological imbalances \citep{dobson1989population}. Instead of focusing solely on eliminating adversarial forces, a more sustainable security model embraces structured engagement with these forces, ensuring that systems evolve in response to emerging threats.

The key insight from biological and digital security ecosystems is that resilience is not about eradicating threats but about learning from them. By integrating ethical hacking, penetration testing, and continuous system monitoring into cybersecurity strategies, we can develop adaptive and responsive security infrastructures. A security framework that thrives under stress, rather than collapses in the face of new challenges, is not only more effective but also essential in an era of ever-evolving digital threats.

\section{Discussion}
\subsection{Summary of Key Insights}
This paper highlights the role of adversarial forces—whether biological or digital—in exposing vulnerabilities and driving system resilience. In both ecosystems, persistent threats compel adaptive responses: hosts evolve immune defenses to counteract parasites \citep{anderson1991infectious}, while cybersecurity frameworks continuously refine encryption protocols, predictive threat modeling, and AI-driven anomaly detection to mitigate evolving cyber threats \citep{shostack2014threatmodeling}. This dynamic mirrors a broader principle of adaptation: the presence of persistent stressors fuels innovation and fortification \citep{schneier2000secrets}.

By reframing adversarial agents as stress-testers of complex systems, this work underscores an essential insight: \textbf{security is not about achieving perfect immunity but about fostering adaptive, responsive resilience}. The concept of controlled exposure, seen in biological vaccination and cybersecurity penetration testing, emphasizes the necessity of structured risk-taking in enhancing system robustness. This aligns with the principle of hormesis—where low-level stressors can strengthen systemic defenses over time \citep{calabrese2019hormesis}. In cybersecurity, this translates to continuous testing, iterative security improvements, and adversarial simulation exercises to harden defenses against real-world attacks.

\subsection{Adaptive Security and Predictive Resilience}
Engagement with adversarial forces has shaped both immune system evolution and cybersecurity strategies. In immunology, repeated exposure to pathogens has driven the emergence of adaptive immune systems, enabling rapid recognition and neutralization of threats before they become lethal \citep{murphy2012janeway}. Similarly, cybersecurity has evolved toward predictive threat modeling, AI-driven intrusion detection, and self-healing network architectures, where systems not only defend against threats but also learn from them to preemptively neutralize future attacks \citep{biggio2018wild}. This proactive approach recognizes that resilience is not about preventing every possible attack, but ensuring that systems can withstand, respond to, and recover from inevitable breaches.

Furthermore, adversarial interactions are not inherently destructive; they are often a necessary catalyst for innovation. Over-securing a system can lead to fragility—just as over-sterilized environments can weaken immune function by reducing exposure to beneficial microbial interactions \citep{dobson1989population}. Similarly, excessive cybersecurity restrictions that discourage ethical hacking and security research may unintentionally leave systems vulnerable to undetected exploits. Instead, a balanced security-first approach embraces structured engagement with adversarial forces, fostering continuous adaptation, innovation, and long-term system sustainability.

\subsection{Implications for Security Strategy}
Understanding adversarial forces as inherent and inevitable components of any system challenges conventional security paradigms that focus solely on eliminating threats. Instead of aiming for absolute protection, security strategies should integrate continuous monitoring, real-time anomaly detection, and adversarial simulation exercises. 

By shifting from a reactive mindset to an adaptive, resilience-driven approach, both biological and cybersecurity ecosystems can develop more robust, self-reinforcing defense mechanisms. The key lesson from parasite-host coevolution and cybersecurity evolution is that resilient systems are not static—they thrive by continuously learning, adapting, and strengthening their defenses against emerging threats.

\subsection{Key Takeaways}
Resilience in both biological and digital systems is not a function of static defense mechanisms but rather an ongoing process of learning, adapting, and evolving in response to stressors. This paper has explored how parasites and hackers—often perceived as adversarial forces—play a fundamental role in driving this process. Instead of being framed solely as threats to be eliminated, they should be understood as forces that compel necessary adaptation, revealing hidden weaknesses and pushing systems toward greater robustness.

The study of biological resilience against parasitic threats offers a compelling analogy for designing adaptive cybersecurity frameworks. In nature, species do not survive by eliminating every parasite; rather, they co-evolve, refining their immune defenses while maintaining ecological balance. Similarly, digital security cannot rely solely on static fortifications—it must embrace continuous adversarial learning, ensuring that systems evolve in tandem with emerging threats.

This shift in mindset extends beyond security professionals—it is a societal imperative. As cybersecurity threats become more sophisticated, society must cultivate a proactive, adaptive approach to digital security. This requires:

\begin{itemize}
    \item \textbf{Policy frameworks that incentivize ethical hacking and controlled security testing.}  
    Cybersecurity must move beyond reactive measures and embrace \textbf{structured adversarial engagement} as a key mechanism for strengthening digital defenses. Ethical hackers play an essential role in exposing systemic vulnerabilities, much like how controlled exposure to pathogens allows biological systems to develop immunity. Governments and regulatory bodies should establish clear legal protections and incentives for ethical hacking, ensuring that security research does not face legal repercussions. Expanding bug bounty programs and mandating \textbf{cybersecurity stress tests} for critical infrastructure—akin to financial sector stress testing—can help organizations assess their resilience against sophisticated cyber threats. Just as the study of parasites informs immune response strategies, \textbf{cybersecurity must integrate adversarial testing as a foundational principle rather than an afterthought}.  

    \item \textbf{AI-driven adaptive security that learns from past threats and anticipates future risks.}  
    Traditional cybersecurity approaches rely on static threat models that struggle to keep pace with \textbf{evolving attack strategies}. Instead, \textbf{adaptive security frameworks} should function more like biological immune systems, learning from previous threats and dynamically adjusting defenses. AI-driven security can continuously monitor for anomalous behavior, applying \textbf{reinforcement learning and adversarial training} to preemptively mitigate emerging risks. Just as immune cells retain memory of past infections, \textbf{AI models should incorporate historical cyberattack data} to refine their response strategies. Moreover, distributed cybersecurity architectures—similar to decentralized immune networks—could enhance \textbf{real-time collaboration between organizations} without compromising sensitive data. This shift toward \textbf{self-learning, self-correcting security mechanisms} is crucial for keeping pace with increasingly sophisticated cyber threats.  

    \item \textbf{Cross-disciplinary research bridging epidemiology and cybersecurity to develop more resilient security models.}  
    Cybersecurity can benefit from frameworks originally designed to track and control \textbf{disease outbreaks}, where early detection, containment, and adaptive response determine resilience. Concepts such as \textbf{infection modeling (R0), mutation tracking, and immune memory} could serve as inspiration for new cyber threat modeling techniques. For instance, just as epidemiologists \textbf{monitor viral mutations} to predict the next pandemic strain, cybersecurity researchers could adopt \textbf{evolutionary threat intelligence}, tracking how malware adapts to different environments and designing countermeasures accordingly. Similarly, the concept of \textbf{herd immunity in public health} suggests new approaches for distributed cybersecurity, where \textbf{shared defense mechanisms reduce systemic risk}. By integrating lessons from epidemiology, cybersecurity researchers can develop \textbf{adaptive, predictive, and decentralized security architectures} that evolve in response to dynamic threats.
\end{itemize}

\subsection{Future Directions}
The intersection of epidemiology and cybersecurity offers a powerful framework for understanding resilience as a dynamic, evolving process rather than a fixed state. Both biological and digital ecosystems are increasingly complex and interconnected, facing threats that are not only persistent but adaptive. To build robust defenses, systems must not aim to eliminate threats entirely but rather integrate mechanisms that enable them to recognize, learn from, and adapt to emerging risks in real time.

One promising research direction is the development of \textbf{AI-driven predictive security models} that borrow from epidemiological forecasting. In public health, predictive models analyze pathogen transmission patterns to anticipate outbreaks and guide interventions \citep{morens2020emerging}. A similar approach in cybersecurity could leverage machine learning-based anomaly detection, where security models continuously refine their defenses by learning from evolving attack patterns—much like immune systems adapt to new pathogens over time \citep{schmid2014antigenic}. By integrating real-time threat intelligence, AI-driven security frameworks could proactively identify and neutralize vulnerabilities before they are exploited.

\textbf{Structured adversarial engagement} is another crucial component of future cybersecurity strategies. Just as vaccine research strategically exposes immune systems to controlled threats, cybersecurity must institutionalize ethical hacking initiatives to fortify system defenses. Red Team vs. Blue Team simulations, penetration testing, and adversarial learning are essential in identifying weaknesses before malicious actors can exploit them \citep{dobson1989population}. However, regulatory policies must strike a balance—overregulating ethical hacking risks suppressing security research, while an open but structured engagement model can incentivize responsible disclosures and continuous improvement in security infrastructure \citep{holt2012hackers}.

A particularly transformative direction is the development of \textbf{self-adaptive digital immune systems}—autonomous, self-correcting security frameworks inspired by biological immune responses. These systems would integrate adaptive learning algorithms that detect anomalies, classify threats, and dynamically adjust security protocols in real time. Just as immune cells "remember" past infections and mount faster responses upon re-exposure, AI-driven security architectures could develop an adaptive threat memory, allowing them to proactively neutralize attacks before they escalate \citep{chang2020modelling}. Research in this area suggests that leveraging adversarial interactions productively may be key to designing security ecosystems that anticipate threats rather than merely react to them.

These interdisciplinary approaches — combining predictive epidemiological modeling, structured adversarial engagement, and digital immune systems — illustrate a broader paradigm shift. Rather than viewing security as a static problem to be solved, it must be treated as an \textbf{ongoing, evolutionary process}. Systems that learn from disruptions instead of merely resisting them will be the most resilient in the face of future threats.

\section{Conclusion}
The study of nature’s battle-tested defense mechanisms provides a powerful roadmap for rethinking security in the digital age. Just as host-pathogen interactions have driven the evolution of increasingly sophisticated immune defenses, the ongoing interplay between adversarial forces and cybersecurity professionals continues to shape the resilience of digital ecosystems. 

True security is not about achieving perfect protection—it is about developing the capacity to \textbf{anticipate, withstand, and evolve in response to threats}. By embracing principles of \textbf{adaptive learning, adversarial engagement, and predictive resilience}, cybersecurity can move beyond static defense models and toward a more dynamic, self-reinforcing framework. 

The future of security does not lie in eliminating adversarial forces but in designing systems that \textbf{thrive under pressure, learn from challenges, and evolve continuously}. In both biological and digital domains, resilience is not just a safeguard—it is an active, ongoing process that defines the survival and strength of complex systems.

\bibliographystyle{plainnat}
\bibliography{references}

\end{document}